\documentclass{ws-p8-50x6-00}
\input{epsf}

\def\bq{\begin{equation}}
\def\eq{\end{equation}}
\def\be{\begin{eqnarray}}
\def\ee{\end{eqnarray}}

\def\prl {Phys. Rev. Lett. }\def\pr{Phys. Rev. }
\def\np{Nucl. Phys. }

\def\roughly#1{\mathrel{\raise.3ex\hbox{$#1$\kern-.75em
\lower1ex\hbox{$\sim$}}}}

\begin{document}

\title{ Orbital Angular Momentum Parton Distributions in Quark Models}

\author{Sergio Scopetta and Vicente Vento}

\address{
Departament de Fisica Te\`orica, Universitat de Val\`encia,
46100 Burjassot (Val\`encia), Spain
\\E-mail: sergio.scopetta@uv.es, vicente.vento@uv.es}

\maketitle

\abstracts{
At the low energy, {\sl hadronic}, scale we calculate 
Orbital Angular Momentum (OAM) twist-two parton distributions
for the
relativistic MIT bag model and for non-relativistic quark models.
We reach the scale of the data by leading order evolution in
perturbative QCD.
We confirm that the contribution of quarks and gluons OAM
to the nucleon spin grows with $Q^2$,
and it can be relevant at the experimental scale,
even if it is negligible at the hadronic scale, irrespective of the model
used.
The sign and shape of the quark OAM distribution 
at high $Q^2$
may depend strongly on the relative size of the OAM and spin
distributions at the hadronic scale. 
Sizeable quark OAM distributions at the hadronic scale, 
as proposed by several authors, can produce 
the dominant contribution to the nucleon spin at high $Q^2$. 
}

\section{Introduction}
Understanding how the partons carry the angular momentum in 
the nucleon has become a main effort of present day physics. The quark spin 
contribution $\Delta \Sigma$ is well defined in QCD \cite{qspin}, 
thus measurable; with respect to the gluon spin contribution $\Delta g$,
the experimental \cite{dw} and theoretical \cite{dra} situation
has been discussed at the Conference.
Our knowledge of 
the quark, $L_q$ and gluon, $L_g$, Orbital Angular Momentum (OAM) 
is less satisfactory.

It is well known, that the most natural definition of 
OAM for quarks and gluons cannot be separated in a gauge invariant way from
the corresponding spin terms \cite{ja-ma}. However, recently, new definitions 
of angular momentum have been implemented to accommodate gauge invariance
\cite{ji-2,ba-ja}, and from them adequate twist two OAM distributions 
have been constructed.

In this respect, three different possibilities have been investigated.
One proceeds by choosing a particular gauge \cite{ja-ma,ji-1,scha-1}. 
The OAM operator leads to the naive definition of orbital angular momentum 
up to effects not controlled by the gauge fixing. A second, 
maintains gauge invariance, by loosing covariance, 
defining the distributions only in the 
class of reference frame where the nucleon has a definite polarization 
\cite{hood}. In this case the resulting distributions can be related to
the forward limit of off-forward quantities, and are measurable  
\cite{ji-2,hood,hood-1}. The last proceeds by 
defining  OAM operators such that the distributions are
gauge invariant \cite{ba-ja} . Furthermore, in
the light-cone gauge, they reduce to the
natural definitions \cite{ja-ma}. At present, however, no physical
process has been proposed to access them.

Evolution equations for the OAM distributions have been derived
\cite{ji-1,scha-1,hood,har,ter}, 
%
and they have been numerically  solved by using as input data-inspired
OAM distributions \cite{scha-2}.

During the last few years we have developed a scheme to study distributions
based on model calculations \cite{tr97}. This procedure has been
worked out thus far to leading order in the twist expansion and 
therefore we should compare with the data only at high $Q^2$,  
where the contribution from the non leading twists vanishes. 
In this talk we present the OAM distributions 
obtained by evolving those, 
properly calculated within relativistic and non relativistic
models, to the experimental scale \cite{scop}.


\section{The theoretical framework}

It has been suggested in the past that the OAM contribution to the nucleon spin
can be large at low \cite{vento,seg}, as well as large \cite{qspin}, energy 
scales. 
Since the quark spin, $\Delta \Sigma(Q^2)$, 
and the gluon spin, $\Delta g(Q^2)$, are 
observables, one may use the Spin Sum Rule
\cite{ja-ma} for the nucleon,

\bq
{1 \over 2} \Delta \Sigma(Q^2) + \Delta g (Q^2)
+ L_q(Q^2) + L_g(Q^2) = {1 \over 2}
\label{sr}
\eq
to determine the global OAM contribution to the spin. 
The problem of
separating 
the OAM in the quark and gluon fractions in a gauge invariant way,
already addressed in \cite{ja-ma}, is a cumbersome one.

We will adopt here the gauge 
independent, twist-2, new definition for the quark OAM
distribution, given in ref.
\cite{hood}.  This definition
mixes the polarized and unpolarized singlet quark distributions
with the forward limit of off-forward parton distributions
(OFFPD).
Since the OAM matrix element
is not a Lorentz scalar, there is an ambiguity in this definition
for any relativistic quantum theory. To avoid it, one  must take for QCD 
a system of coordinates where the nucleon has a definite helicity, 
thus loosing covariance. 

We proceed to use this last theoretical development, 
to perform a phenomenological model analysis of the OAM distributions, 
using relativistic as well as non-relativistic quark models. 

Let us first discuss the definition of the OAM for a relativistic model.
According to Ref. \cite{hood},
if we  assume 
that the nucleon is moving in the $z$ direction and is polarized 
with helicity $+1/2$ , the
quark OAM distribution is given by,

\bq
L_q(x,Q^2) = {1 \over 2} \left[ x(\Sigma(x,Q^2) + E_q(x,Q^2)) -
\Delta \Sigma(x,Q^2) \right],
\label{lqr}
\eq
where $\Sigma(x,Q^2)$ $(\Delta \Sigma(x,Q^2))$ is the usual  
unpolarized (polarized) $singlet$ quark distribution
and $E_q(x,Q^2)$ is the $forward$ $limit$ of the
helicity-flip, chiral {odd}, twist-two OFFPD 
$E(x,\Delta^2,\Delta \cdot n)$ \cite{ji-2}.
The latter quantity is defined through the twist-2 part
in the the following twist expansion

\begin{eqnarray}
\int {d \lambda \over 2 \pi} \langle P' |
\bar \psi(- \lambda n /2) & \gamma^\mu & \psi(\lambda n/2) |P \rangle
=   
H(x, \Delta^2,\Delta \cdot n) \bar U(P') \gamma^\mu U(P)
\nonumber \\
& + & E(x, \Delta^2,\Delta \cdot n) \bar U(P') 
{i \sigma^{\mu\nu} \Delta_\nu \over 2 M} U(P)+ ...,
\label{corr}
\end{eqnarray}
where $P(P')$ is the 4-momentum of the initial (final) nucleon
in a virtual Compton scattering process, $\Delta_\mu=P'_\mu - P_\mu$,
$n = (1,0,0,-1)/2\Lambda$, $\Lambda$ is fixed 
by the choice of the reference frame, 
and $H(x, \Delta^2,\Delta \cdot n)$ is the helicity-conserving,
chiral even, twist-2
OFFPD whose forward limit is the usual forward
unpolarized parton distribution.
In the definitions above, the forward limit corresponds to
$\Delta^2 \rightarrow 0$, $\Delta \cdot n \rightarrow 0$.

In the non-relativistic case, the nucleon
wave function is given in general by an expansion in terms of the 
eigenstates of
some approximate hamiltonian. Let N label the quantum numbers of the
eigenstates, i.e., principal quantum number, 
orbital angular momentum, spin, ..., then

\bq
\Psi(\vec p_1, \vec p_2, \vec p_3) = 
\sum_{N} a_{N} 
\psi_{N} (\vec p_1,\vec p_2,\vec p_3)
\label{calL}
\eq
The OAM quark parton distribution, generalizing our
approach developed for unpolarized and polarized distributions
\cite{tr97}, is determined by \cite{scop}

\bq
L_q(x,Q^2) = 2 \pi M \sum_{N} 
|a_{N}|^2 \int_{|p_-(x)|}^\infty 
\, dp \, p \, L_{q,{N}}^z(p)~,
\label{lqnr}
\eq
where

\bq
L_{q,{N}}^{z}(p) = 
\langle \psi_{N} (\vec p_1,\vec p_2,\vec p_3)|
\sum_{i=1}^3 L_i^{z} \delta (\vec p - \vec p_i)
|\psi_{N} (\vec p_1,\vec p_2,\vec p_3) \rangle,
\label{lqz}
\eq
and the lower integration limit is fixed by energy conservation
\cite{tr97}.
%

Having set down the framework which defines the OAM distributions 
at the hadronic scale, we proceed to calculate them explicitly 
in two models and
to study their evolution.

\section{OAM parton distributions at high $Q^2$}

We proceed to study the OAM distribution in two different scenarios for proton
structure: i) a non-relativistic scheme based on the Isgur-Karl model 
\cite{ik78};
ii) a relativistic approach, as described by the MIT bag model \cite{ja74}.

For the non-relativistic scenario we consider initially
the Isgur-Karl model with a proton wave function given by a harmonic oscillator
potential 
including contributions up to the $2 \hbar \omega$ shell
\cite{gia}. In this case  the wave function, Eq. (\ref{calL}), is given by 

\bq
|N \rangle = 
a_{\cal S} | ^2 S_{1/2} \rangle_S +
a_{\cal S'} | ^2 S'_{1/2} \rangle_{S} +
a_{\cal M} | ^2 S_{1/2} \rangle_M +
a_{\cal D} | ^4 D_{1/2} \rangle_M~,
\label{ikwf}
\eq
where we have used the spectroscopic notation $|^{2S+1}X_J \rangle_t$, 
with $t=A,M,S$ being the symmetry type.
The coefficients were determined by spectroscopic properties to be: 
$a_{\cal S} = 0.931$, 
$a_{\cal S'} = -0.274$,
$a_{\cal M} = -0.233$, $a_{\cal D} = -0.067$.

Calculating Eq. (\ref{lqz}) using the wave function Eq. (\ref{ikwf}),
one gets for the parton distribution, Eq. (\ref{lqnr})\cite{scop},

\bq
L_q(x,\mu_0^2) = |a_{\cal D}|^2 { M \over \alpha \sqrt{\pi}}
\left ( {3 \over 2} \right )^{3/2} 
\left( 
{1 \over 5} {p_-^4 \over \alpha^4} + 
{13 \over 30} {p_-^2 \over \alpha^2} + 
{23 \over 45} 
\right) 
e^{ - { 3 p_-^2 \over 2 \alpha^2} }~,
\label{lqik}
\eq
where $\alpha^2$ is a  parameter of the model
.
Note that only the small $|^4 D_{1/2}\rangle_M$ 
wave component gives a contribution to the distribution.

In the relativistic scenario, we evaluate 
the twist-two distribution Eq. (\ref{lqr})
in 
the MIT bag model.
The term $\Delta \Sigma(x,\mu_0^2)$ in Eq. (\ref{lqr})
is discussed for the bag
in ref. \cite{ja-ji}.
To calculate Eq.(\ref{lqr}) we need the quantity
$\Sigma(x,\mu_0^2)+E_q(x,\mu_0^2)$, which is given by the forward
limit
of the OFFPD distribution 
$H(x,\Delta^2,\Delta \cdot n)+
E(x,\Delta^2,\Delta \cdot n)$. The result for the latter in the bag, 
calculated by  using  the general definition (Eq. (\ref{corr})), is
to be found in Eq.(29) of ref. \cite{meln}, 
from which we have obtained 
the forward limit. The explicit expression for such a limit
can be found in Ref. \cite{scop}.

\begin{figure}[t]
\vspace{5cm}
\includegraphics{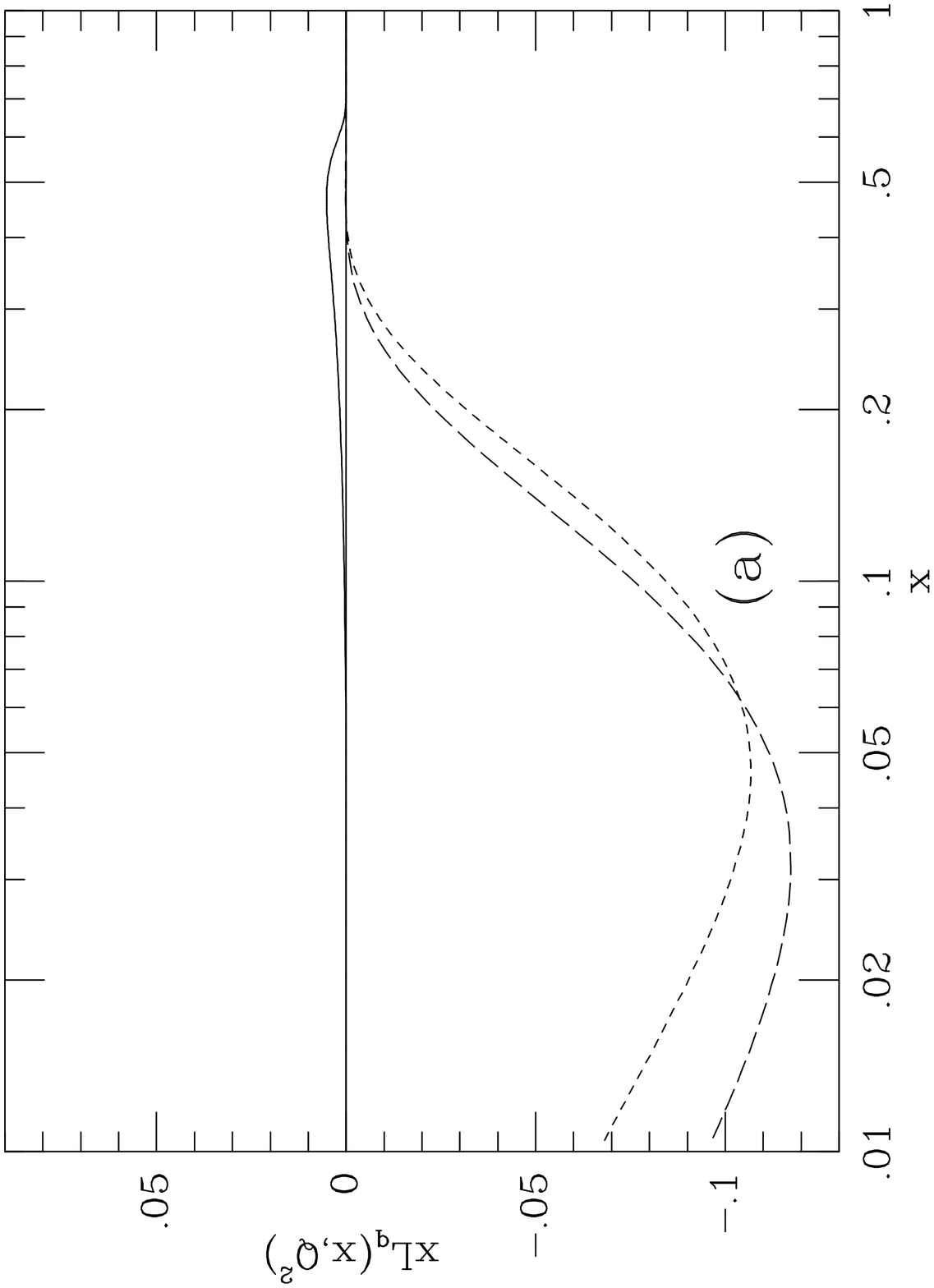}
\includegraphics{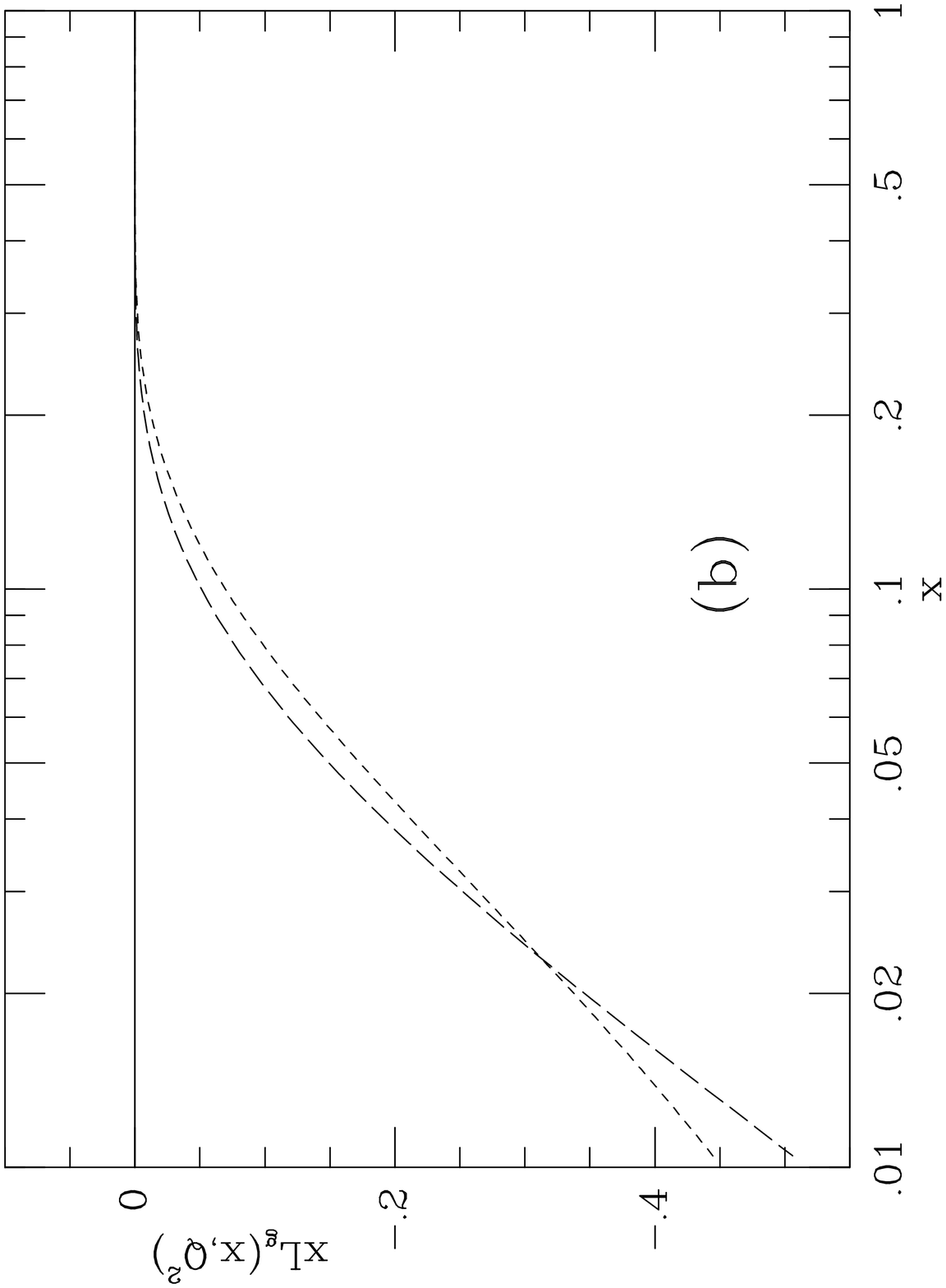}
\caption{ Proton OAM distributions in the IK model for the quarks, (a), 
and for the
gluons, (b).
The full curves show the initial distributions
at the hadronic scale of $\mu_0^2=0.08$ $GeV^2$, where
a negligible fraction of the nucleon momentum is carried by the gluons; 
the dashed curves represent the LO--evolved distributions
at $Q^2=10$ $GeV^2$; the long-dashed curves give the
LO--evolved distributions at $Q^2=1000$ $GeV^2$. }
\end{figure}
In both cases, Isgur-Karl (IK) and MIT,
we use the corresponding support correction 
as defined in \cite{tr97} and \cite{jr80}, respectively.

Once the distributions $L_q(x,\mu_0^2)$
have been obtained at the low (hadronic) scale 
of the model, we perform a LO QCD evolution according 
to the equations displayed in 
refs. \cite{scha-1,hood,har,ter,scha-2}.
These contain a complicate mixing between 
$L_{q(g)}(x,\mu_0^2)$, $\Delta \Sigma(x,\mu_0^2)$
and $\Delta g(x,\mu_0^2)$, a feature 
which will be very relevant in the analysis
of the data.

The results of our analysis are shown in Figs. 1 through 5.
In Fig. 1 we show the  IK result for quarks (gluons) in (a) ((b));
The full curve corresponds to the  initial distribution, which is missing 
in (b), since we start from a very low hadronic scale where no valence 
gluons exist. At LO the hadronic scale corresponds to $\mu_0^2\simeq0.08$ 
$GeV^2$. The dashed curves correspond to the result of the evolution from the
hadronic OAM distributions, to $Q^2=10$ $GeV^2$ (short-dashed) and to
 $Q^2=1000$ $GeV^2$ (long-dashed).

We can summarize the results of the calculation as follows:
i) the evolved distributions are negative;
ii) the magnitude of the distributions increases with $Q^2$
at low $x$; iii) the magnitude of $L_q(x,Q^2)$ is {\sl small} but increases
with respect to the tiny starting distribution
(Eq. (\ref{lqik})); iv) though gluons are assumed
to be negligible at the hadronic scale, $L_g(x,Q^2)$ becomes much larger
than $L_q(x,Q^2)$ at high $Q^2$.
\begin{figure}[t]
\vspace{5cm}
\includegraphics{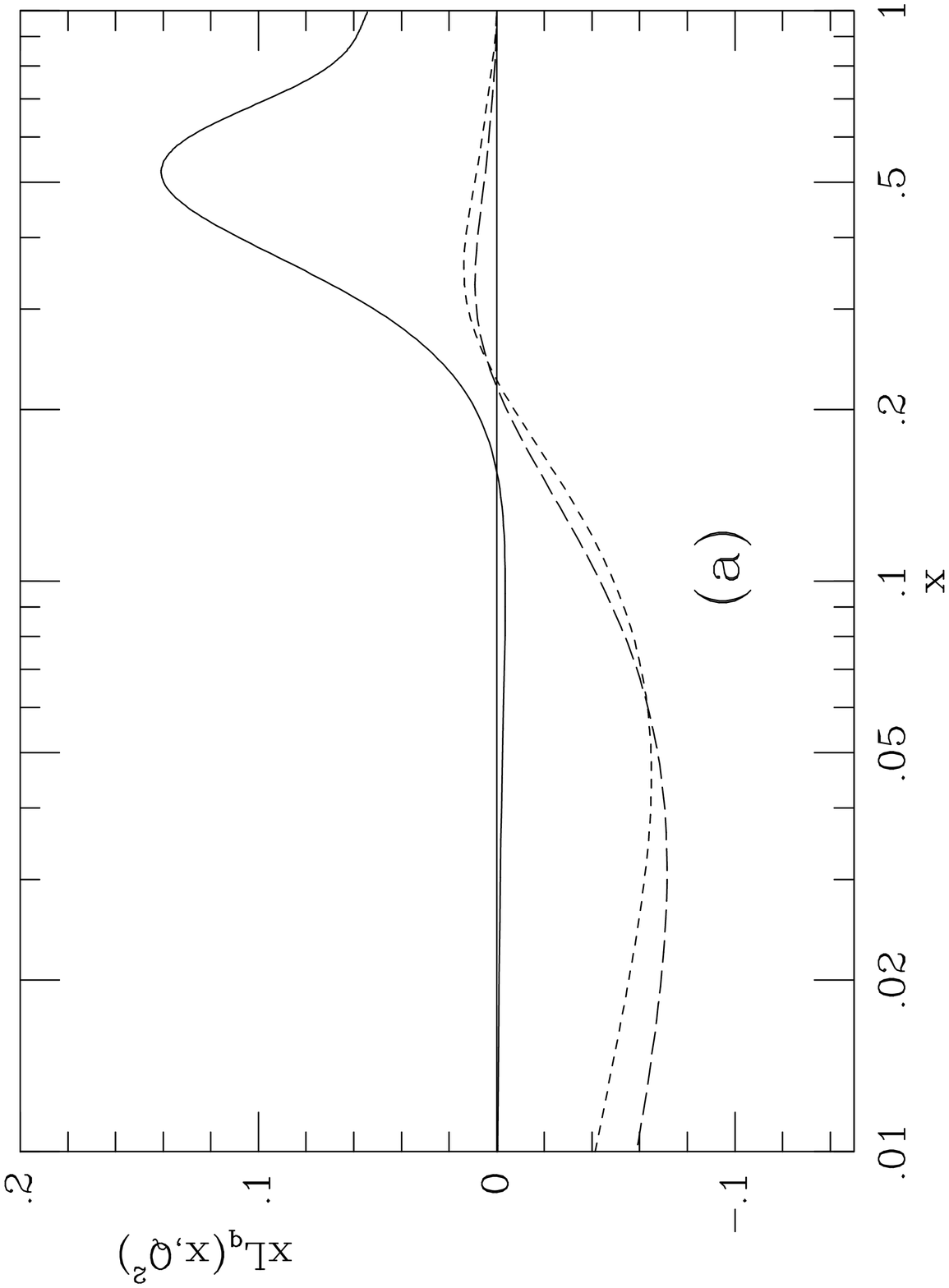}
\includegraphics{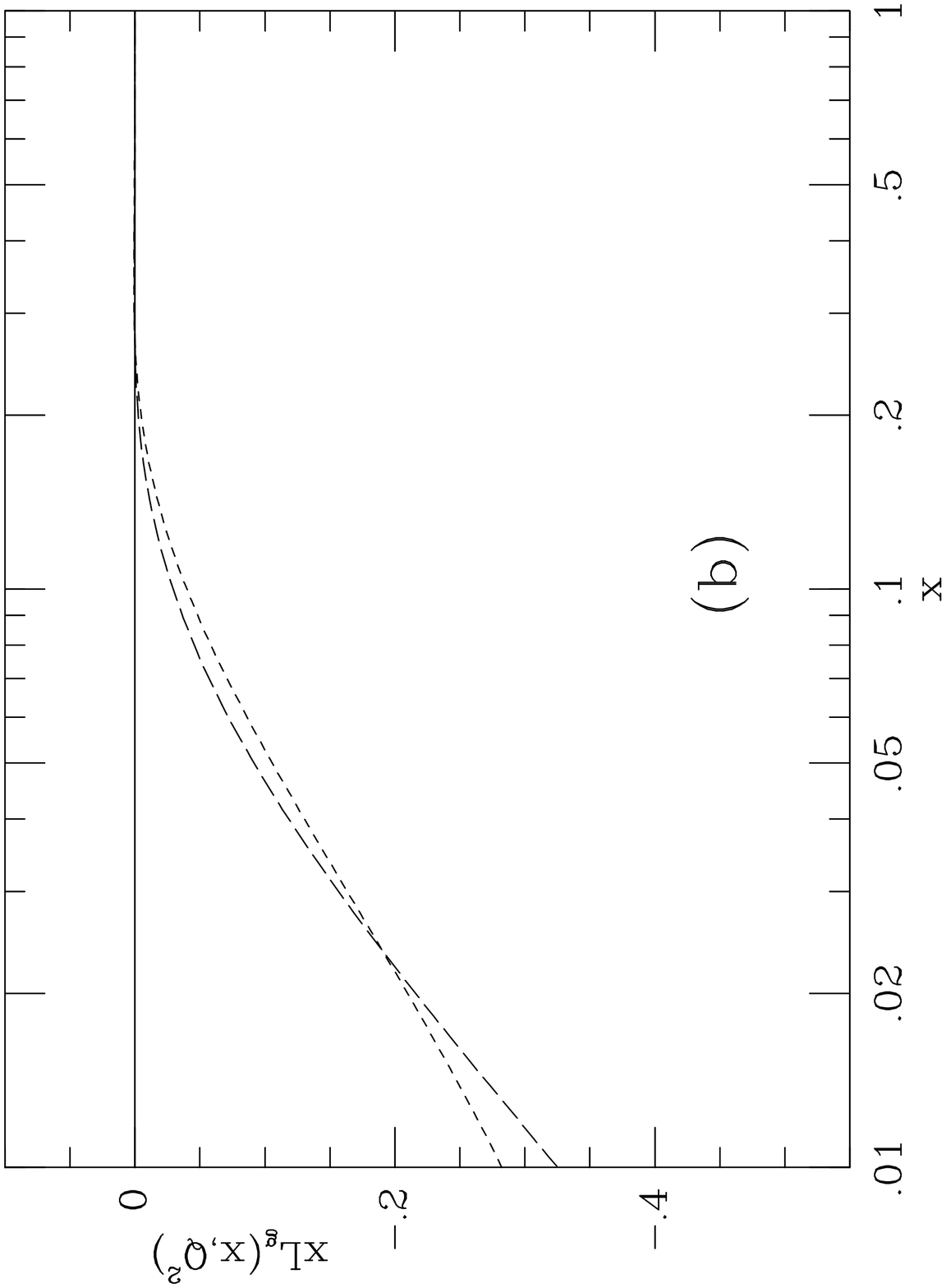}
\caption{ 
The same as in Fig.1, but for the MIT model.
It should be noticed that the distribution 
at the hadronic scale of $\mu_0^2=0.08$ $GeV^2$
is shown as it comes
out from the model calculation, before the support correction is
implemented.
For this reason it does not go to zero at $x=1$.
}
\end{figure}
We show in Fig. 2 the same analysis for the MIT bag model.
Here the initial $L_q$ is much larger, but the result of the evolution is
qualitatively basically the same.

Thus our first conclusion is that there is little model dependence for 
 different
initial OAM distributions. In this case their structure does not seem to 
influence very much the evolution. It is clear that the other inputs of the 
equations, $\Delta \Sigma (x,\mu_0^2)$ and $\Delta  g (x,\mu_0^2)$,
are the dominating features.

For these results the initial scale has been very low and therefore
no initial gluon distribution was required.
In Fig. 3 we show the result of the evolution for the IK model
starting from a higher scale, $\mu_0^2\simeq 0.23$ $GeV^2$.
At this hadronic scale, for LO, about 40 $\%$ of the proton  momentum must be 
carried by the gluons.
The polarized gluon contribution is built starting from
the valence quark distributions as done in \cite{tr97}
and suggested in \cite{reya}, and we define $L_g(x,\mu_0^2)$ from
$L_q(x,\mu_0^2)$ using the same prescription.

Again the same
features as in the first analysis are found. 
\begin{figure}[t]
\vspace{5cm}
\includegraphics{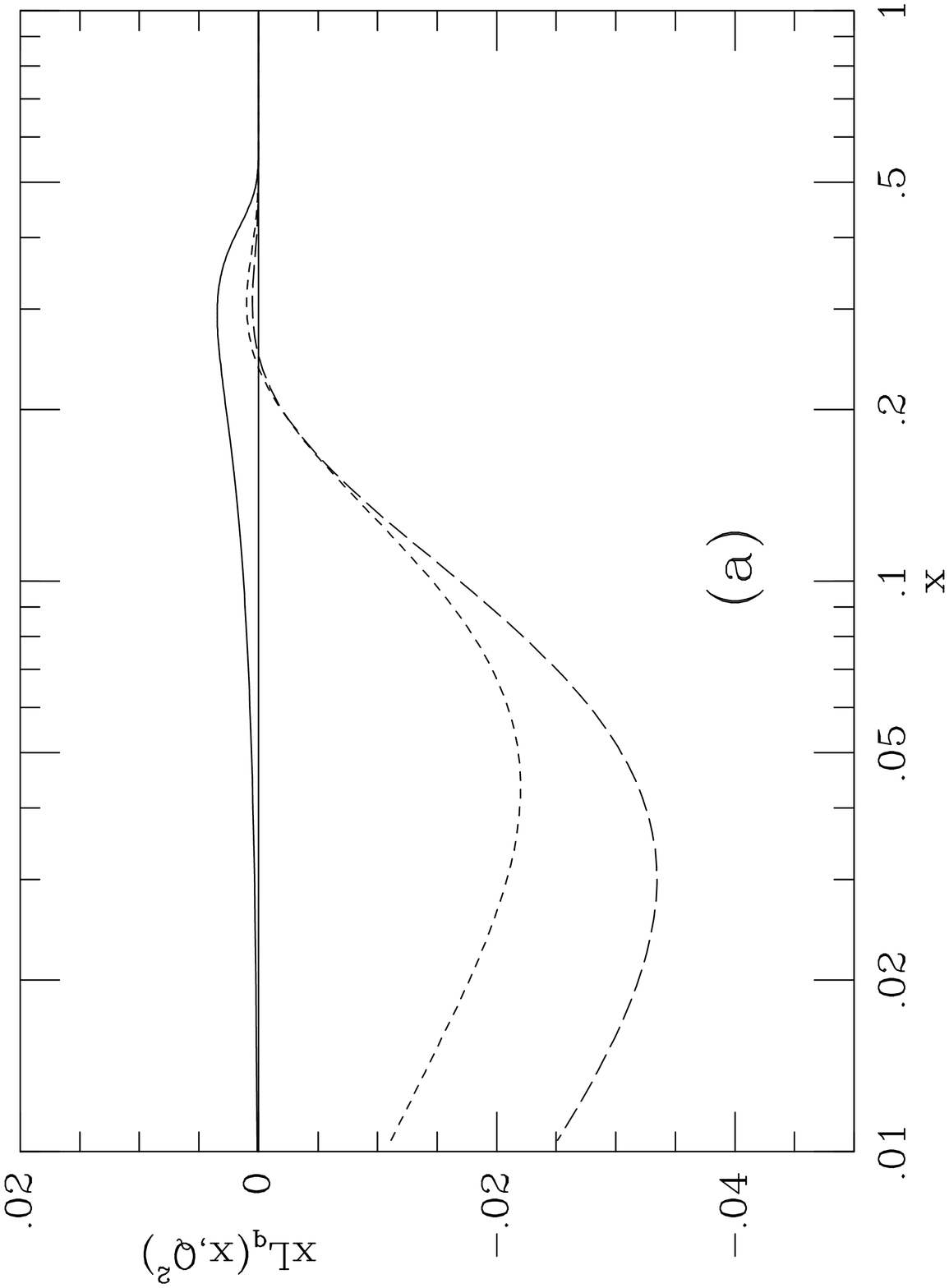}
\includegraphics{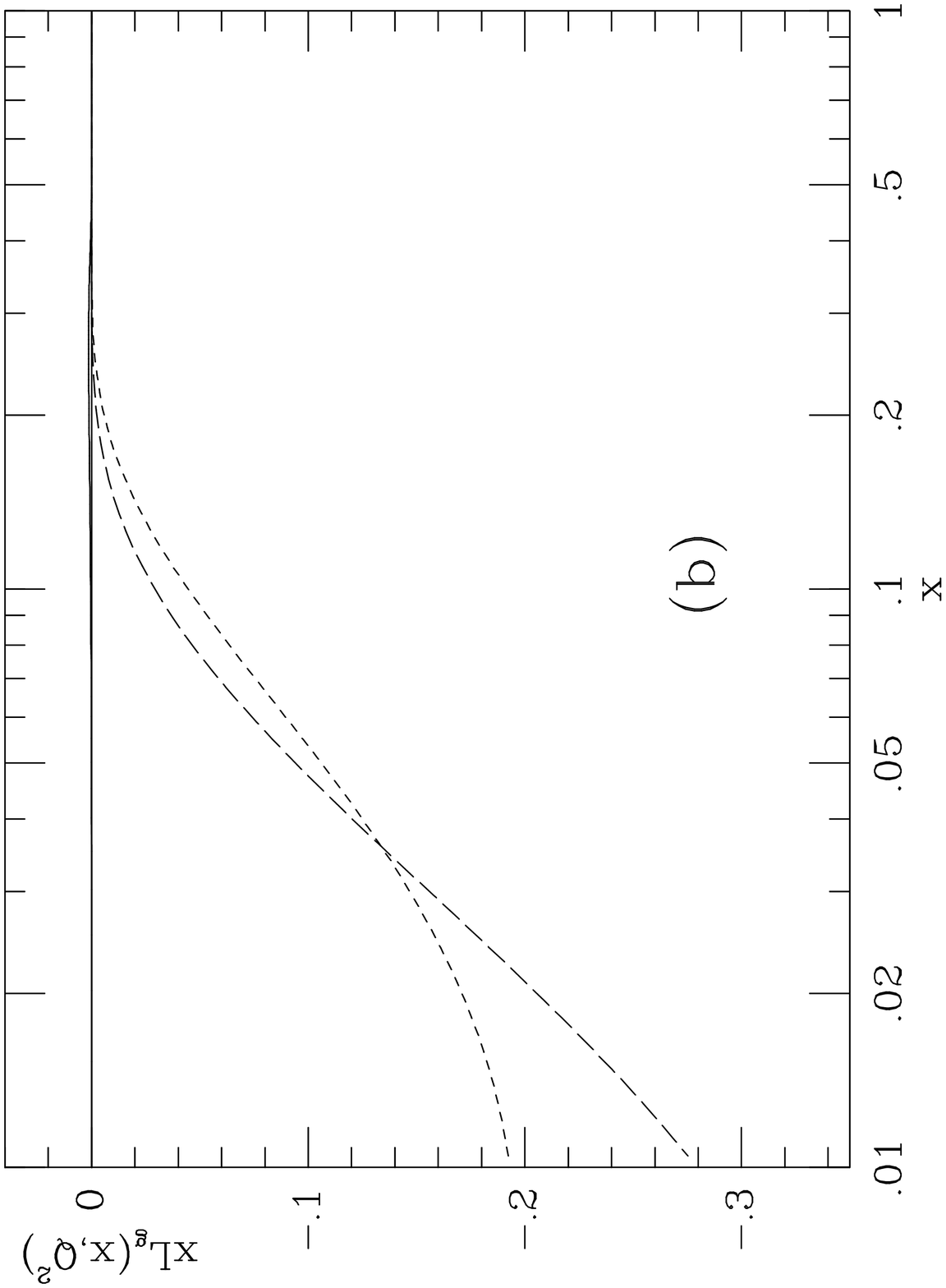}
\caption{
Proton OAM distributions in a modified IK model (see text)
for the quarks, (a), and the gluons, (b).
The full curves show the initial distributions
at the hadronic scale of $\mu_0^2=0.23$ $GeV^2$, where
around 40 \% of the nucleon momentum is carried by the gluons;
the dashed curves represent the LO--evolved distributions
at $Q^2=10$ $GeV^2$; the long-dashed curves give the 
LO--evolved distributions
at $Q^2=1000$ $GeV^2$.
The initial
gluon distribution is so small, in relation with the final one, that
it does not show up in (b).
}
\end{figure}
Recapitulating, our results show that the input distributions 
$L_q(x,\mu_0^2)$ and $L_g(x,\mu_0^2)$
do not seem to determine the behavior of their evolved 
ones, which turns out to be governed by the 
singlet
polarized distributions $\Delta g (x,\mu_0^2)$ and 
$\Delta \Sigma(x,\mu_0^2)$, due to their mixing in the evolution 
equations. To check to what extent such a statement, stressed also in
\cite{scha-2}, is valid,
we analyze in Fig. 4 the results of 
a modification of the IK model, the so called  D model,
already studied in \cite{ropele}.
In this variant model,
the $D$ wave probability is set large to reproduce
the axial coupling constant of the nucleon \cite{vento}. This condition
requires the following choice
of the parameters in Eq. (\ref{ikwf}):
$a_{\cal S} = 0.894$, $a_{\cal S'} = 0$,
$a_{\cal M} = 0$, $a_{\cal D} = -0.447$,
i.e., the probability to find a nucleon in the $D$ wave
is about 20 $\%$.  Moreover we have introduced also in the D Model
scenario polarized valence gluons, as we did for the IK scenario of Fig.3.
From the figure it is clear that, while the result for the gluons 
does not differ in a relevant way from the ones found before 
with the various models, the result for $L_q(x,Q^2)$ does, this distribution 
becoming rather large and positive for large $x$. It is important to
stress that in order for this to occur two mechanisms were needed, a
large initial OAM distribution (as in the MIT bag Model), and a higher
hadronic scale (as provided by the valence gluons). As shown
previously, the independent action of the two mechanisms does not lead
to this behavior. 
\begin{figure}[t]
\vspace{5cm}
\includegraphics{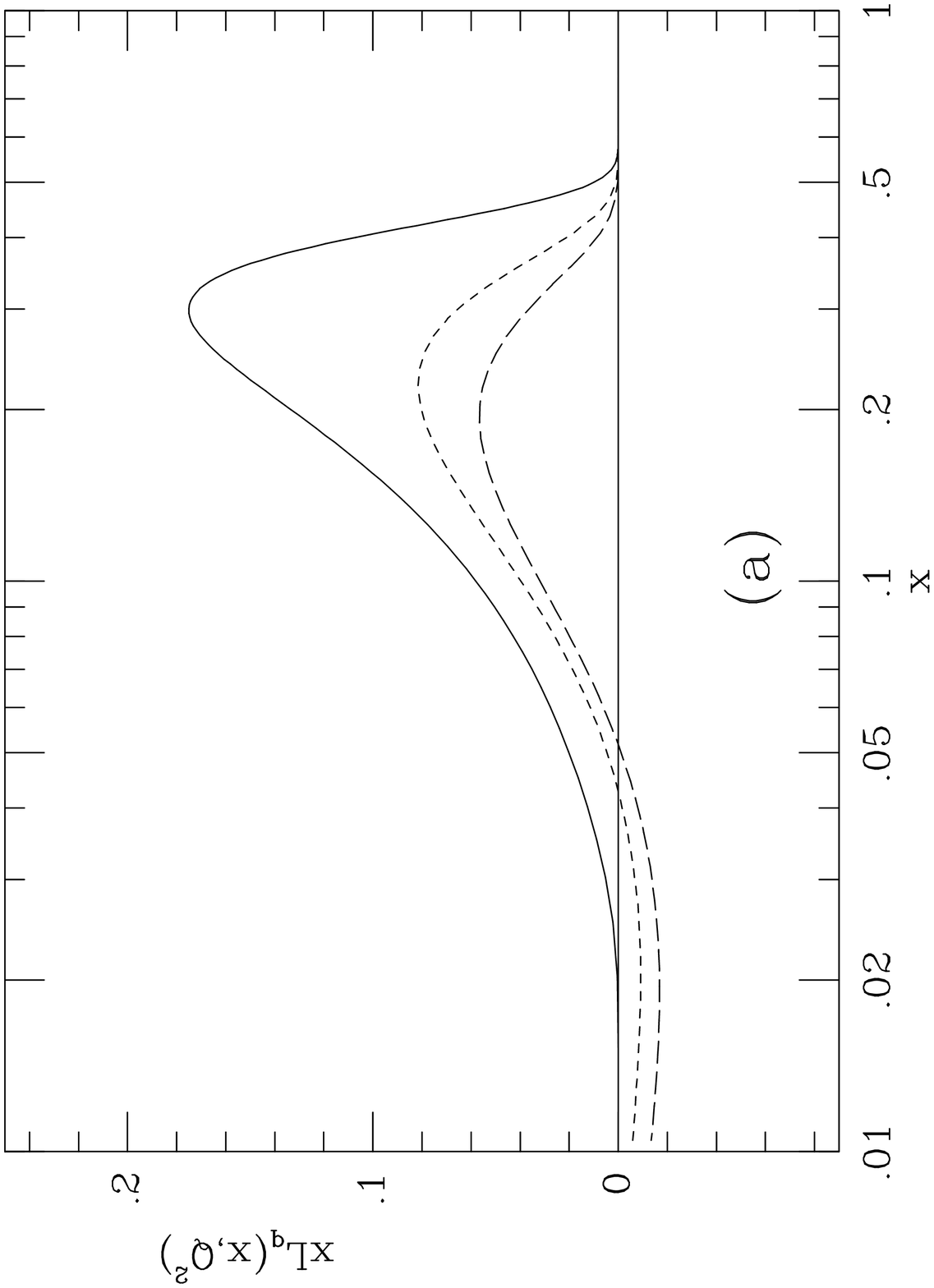}
\includegraphics{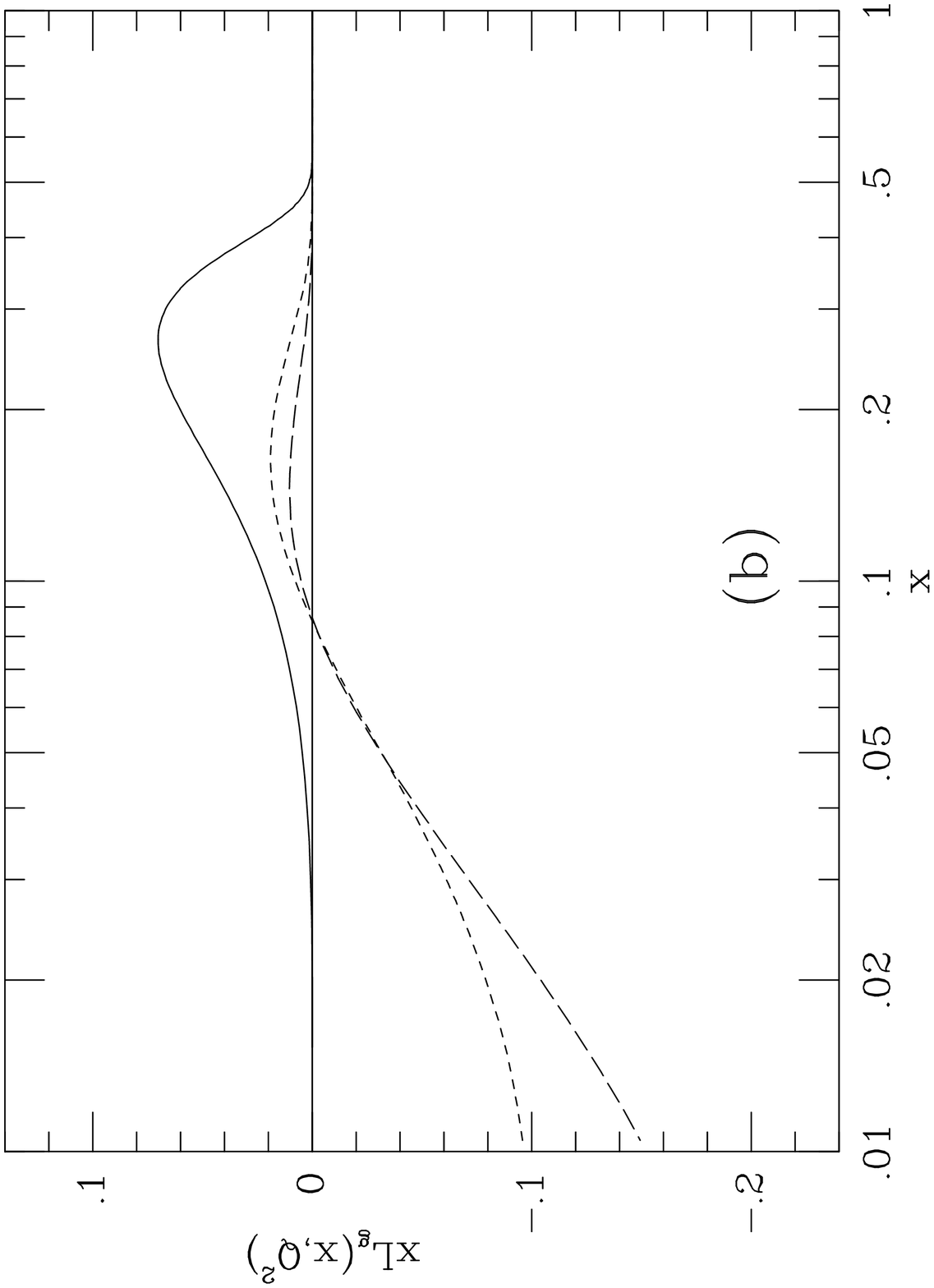}
\caption{
Proton OAM distributions for the ``D-Model'' (see text),
for the quarks, (a), and the gluons, (b).
The full curves show the initial distributions
at the hadronic scale of $\mu_0^2=0.23$ $GeV^2$, where
around 40 \% of the nucleon momentum is carried by the gluons;
the dashed curves represent the LO--evolved distributions
at $Q^2=10$ $GeV^2$; the long-dashed curves give the
LO--evolved distributions at $Q^2=1000$ $GeV^2$.
}
\end{figure}
Thus, $\Delta g (x,\mu_0^2)$ and 
$\Delta \Sigma(x,\mu_0^2)$ are governing
the evolution  as long as they are much larger than
$L_q(x,\mu_0^2)$ at the initial scale. When they have similar size
the above statement is not true any more.
Note that the IK interaction together
with the choice of parameters of the  D model does not
describe the hadron spectrum. Nonetheless, other models
of interaction (for example \cite{vento,seg}) predict at the low scale 
a large OAM and fit the spectrum. 
We conclude therefore that a precise knowledge of the OAM distributions 
will serve to distinguish among the models.
In Fig. 5 we show the evolution of the various 
contributions to the spin sum rule, Eq. (\ref{sr}).
Fig. 5 (a) corresponds to the modified IK scenario
used in Fig. 3, where at the scale of the model
the OAM carried by quarks 
and gluons is very small and the rest of the proton spin
is almost equally shared between quarks and gluons spins
($L_q(\mu_0^2) + L_g(\mu_0^2)\simeq 0.01$,
$\Delta \Sigma(\mu_0^2)\simeq 0.48 $ and
$\Delta g(\mu_0^2)\simeq 0.25 $),
whereas Fig. 5 (b) corresponds to the extreme
scenario used already in Fig 4, the so called  D model,
with a large initial OAM
($L_q(\mu_0^2)=0.145$, $L_g(\mu_0^2)= 0.055$,
$\Delta \Sigma(\mu_0^2)\simeq 0.4 $ and
$\Delta g(\mu_0^2)\simeq 0.1 $).  As predicted by total angular 
momentum conservation \cite{qspin} and already obtained
in \cite{scha-2} as a model-independent feature
of the evolution equation , it is seen that at large $Q^2$
the huge negative contribution $L_g(Q^2)$ basically
cancels out with the positive $\Delta g(Q^2)$.
Anyway, the role of the quark OAM is found to be very
important in the second scenario (cf. Fig. 5(b)),
being at large $Q^2$ the largest contribution to the saturation
of the spin sum rule.
\begin{figure}[h]
\vspace{5cm}
\includegraphics{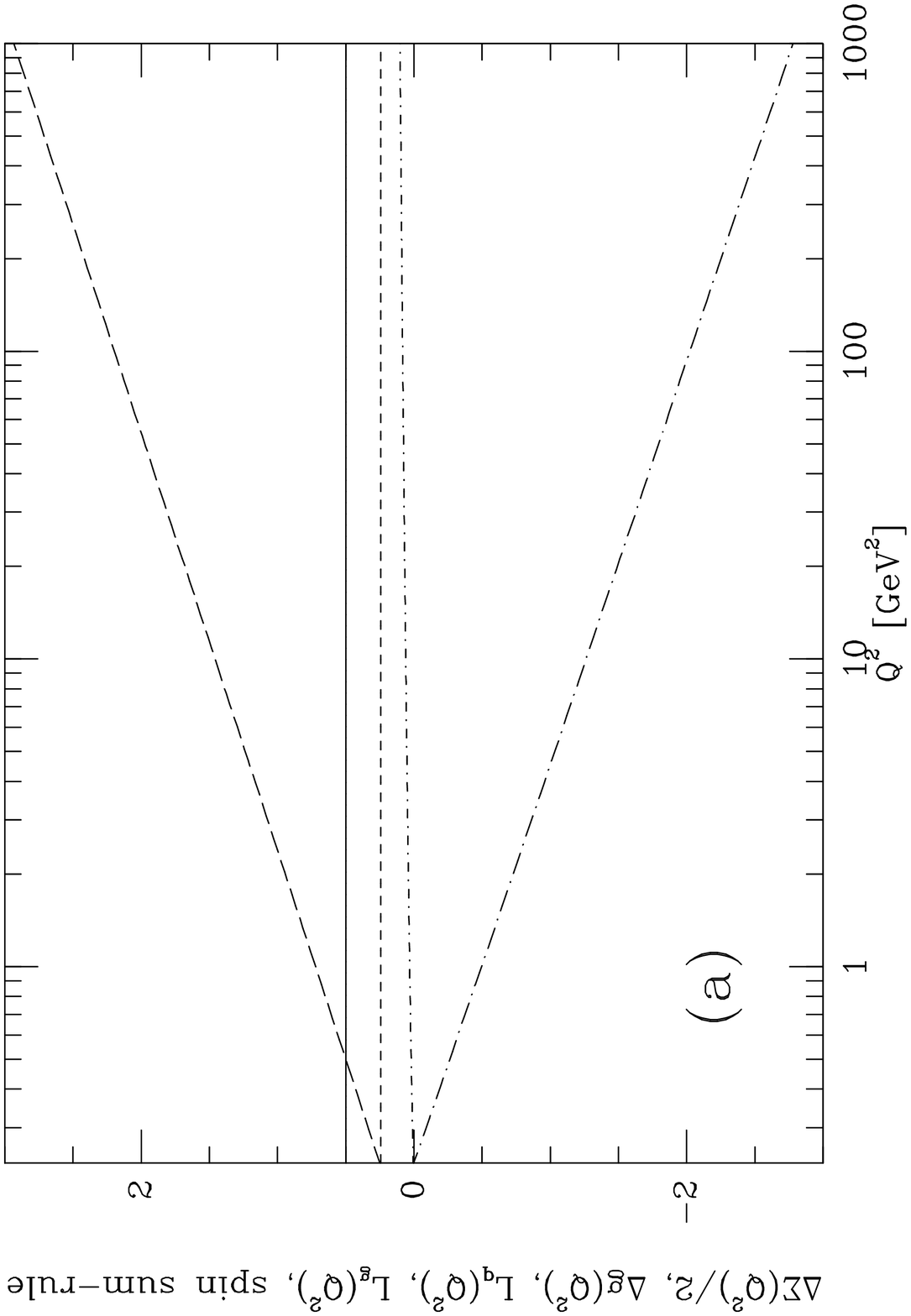}
\includegraphics{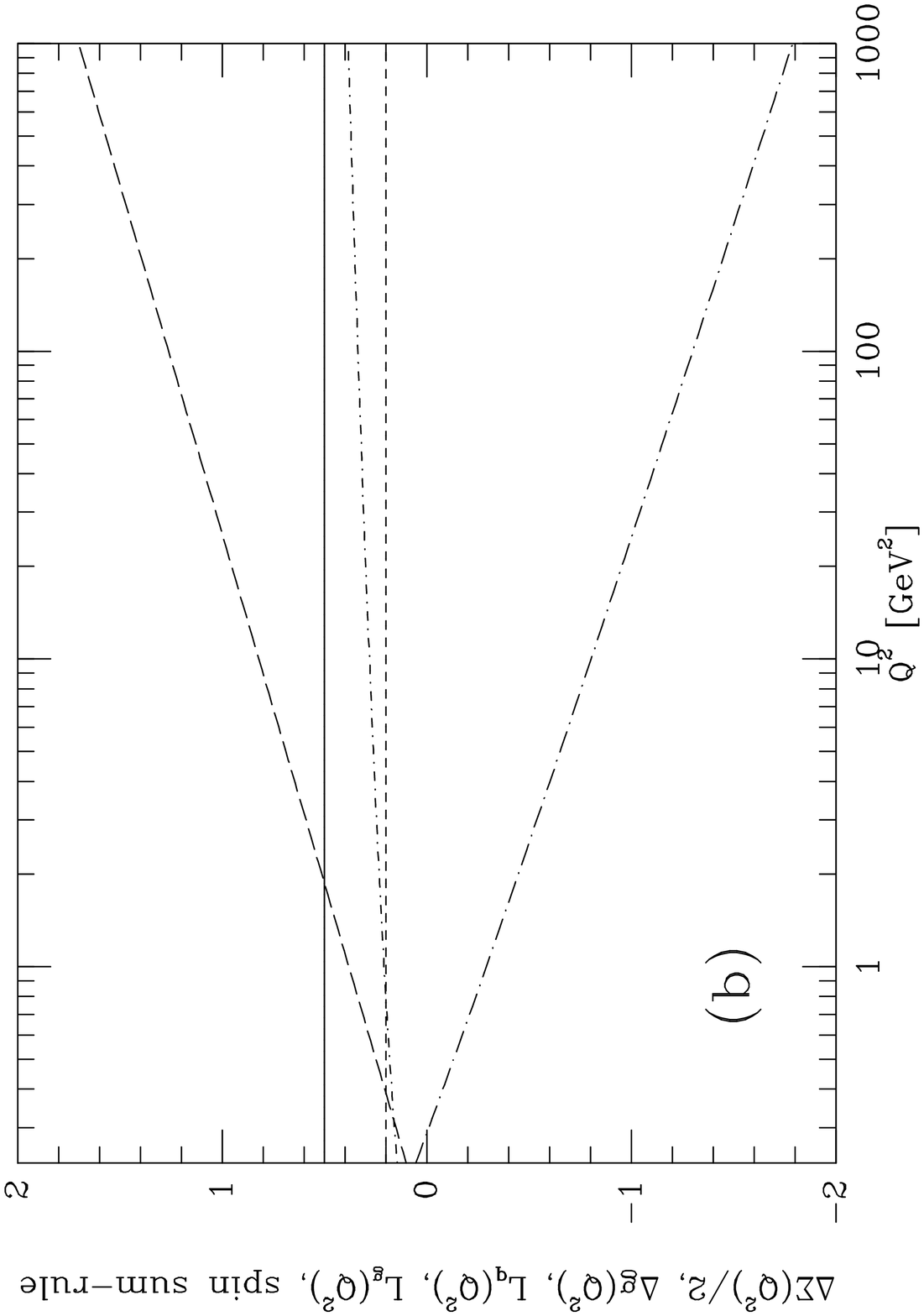}
\caption{
The contributions to the proton spin sum rule,
Eq. (\ref{sr}), according to: (a) the modified IK scenario
of Fig. 3; (b) the ``D model'' scenario of 
Fig. 4. The dashed curve shows ${1 \over 2}\Delta \Sigma(Q^2)$,
the long-dashed one $\Delta g (Q^2)$, the dot-dashed curve is $L_q(Q^2)$,
the dot-long-dashed curve gives $L_g(Q^2)$ and the full curve 
represents the sum of the 
previous four terms, giving the spin sum rule
($J={1 \over 2}$).
}
\end{figure}
Again, we see that quark OAM, due to evolution, can
be important at large $Q^2$ if it is not negligible at the
scale of the model, whereas the gluons OAM, though
it is large, is basically cancelled by the gluons spin. 

\section{Conclusions}

We have studied the OAM distributions as defined newly in order to take into
account gauge invariance. 


We have seen that evolution, as in previous 
calculations\cite{tr97}, plays a major role 
in the outcome of the predictions. The fact that the gluon
and quark spin singlet
distributions mix in these equations with the quark
OAM distributions, implies that
for large $Q^2$ large contributions from the OAM are to be expected, even if
they are not present at the hadronic scale. Thus, two scenarios arise in a
natural way. One, the more conventional one, as described by the more
traditional models, is defined by quark OAM distributions at a small
hadronic scale.
In this scenario the evolved distributions are large, negative and 
almost model
independent and angular momentum DIS physics is dominated by the quark and
gluon spin singlet distributions, 
not by OAM distributions at the hadronic scale. 
A second scenario is defined by quark and gluon OAM distributions at a higher 
hadronic scale. In the latter, soft evolution scenario,
the initial
distributions are important and therefore DIS 
physics may be able to discriminate between models. If the
OAM distributions are large the outcome of the evolution is
strongly dependent on the initial distributions and completely
different from that of the first scenario.

Finally the gluon OAM distributions become huge through evolution, 
even if they are
not present at the hadronic scale. 
However, as it is well known 
\cite{qspin} , 
the gluon OAM and gluon spin
contributions cancel to a great extent in the nucleon spin, 
but not so in other
moments.
Our past experience suggests that LO results provide a 
reasonable {\sl qualitative} 
approximation and we do not expect that NLO corrections
can spoil their general features.

Many phenomenological implications have arisen of our study.
A careful analysis of gauge invariance \cite{ji-2,ba-ja,hood} 
has permitted us to obtain
many observables, which may not only lead to a better understanding 
of the proton
spin, but to describe the proper behavior of QCD at low energies, i.e.,  in the
confining region.
These observations are instrumental in defining  
the picture of the proton 
that should be used for describing low energy properties.
  
\section*{Acknowledgments}

S.S. thanks the organizers for the invitation.
Supported in part by DGICYT-PB97-1227 
and TMR programme of the European Commission ERB FMRX-CT96-008.

\end{document}